\documentclass[12pt,a4paper,final]{iopart}
\expandafter\let\csname equation*\endcsname\relax
\expandafter\let\csname endequation*\endcsname\relax
\usepackage{iopams}
\usepackage{graphicx}
\usepackage{subfigure}
\usepackage[breaklinks=true,colorlinks=true,linkcolor=blue,urlcolor=blue,citecolor=blue]{hyperref}
\usepackage{cite}
\usepackage{braket}
\usepackage{mathtools}

\usepackage[a4paper]{geometry}
\usepackage{bpchem,upgreek}
	\usepackage{amsmath}
	\usepackage{makeidx}
	\usepackage{amsfonts}
	\usepackage[ansinew]{inputenc}
	\usepackage[usenames,dvipsnames]{pstricks}
    \usepackage{graphicx}
    \usepackage{float}
    \usepackage{subfigure}
	\usepackage{pst-grad} 
	\usepackage{pst-plot} 
	\usepackage{sidecap}
	\usepackage[makeroom]{cancel}

\begin{document}

\title[Towards quantum repeater protocol...]{Towards quantum repeater protocol based on the coherent state approach}

\author{M Ghasemi$^{1}$}

\author{M K Tavassoly$^{1}$}
\address{$^1$Atomic and Molecular Group, Faculty of Physics, Yazd University, Yazd  89195-741, Iran}
\ead{mktavassoly@yazd.ac.ir}

\vspace{10pt}
\date{today}

\begin{abstract}
 The aim of this paper is to swap the entanglement between two separate long distant locations. The well-known entangled coherent states as two-mode continuous-variable states are very interesting in quantum teleportation and entanglement swapping processes. To make our investigation more realistic, by using such entangled states as the building block of our quantum repeater protocol, the effect of decoherence on the swapped entanglement is also considered. We explicitly establish our model for four locations, moreover, we find that our model can be extended to $2^N$ locations, where $N=3,4,\cdots$. Consequently, we could introduce this model as a quantum repeater which is helpful for entanglement swapping to enough long distances.
\end{abstract}


\vspace{2pc}

\section{Introduction}
The distribution of entangled states over long distances is an important point in quantum communications. However, due to unavoidable the photon loss in optical channels that leads to entanglement attenuation, transferring the entangled states is interrupted. Using the quantum repeater protocol is a useful way to overcome photon losses \cite{Briegel1998,Duan2001,Zhou2015,Kaiser2013}. Producing and swapping the entanglement are two important basis of quantum repeater protocol. In this protocol long distances are usually divided into several short entangled parts where the states associated with these parts are separable from each other. These parts are then become entangled by the well-known entanglement swapping process. The entanglement between the separated parts can be produced using the beam splitter \cite{Agarwal2013,Pakniat2017,Kim2002,Berrada2009}. In another way, entanglement of these produced entangled states can be swapped by Bell state measurement (BSM) method \cite{Liao2011,Ghasemi2016,Ghasemi2017} and performing interaction \cite{Tacsgiotan2010,Abdel2008,Bashkirov2005}.\\
The quantum repeater protocol has been extended the distance of communication to the order of km ($10^3\sim 10^6$ km) \cite{Jiang2009}. This protocol has been investigated based on Rydberg atomic ensembles \cite{Zhao2010,Han2010} and quantum dots \cite{Wang2012,Wang2014,Predojevic2015}. The creation of entangled pairs which are used in quantum repeater applications has been considered in \cite{Munro2008} in which optimizing the noise properties of the initially distributed pairs significantly improves the rate of generating long-distance Bell pairs. In \cite{Simon2007} the quantum repeaters with photon pair sources have been considered, experimentally. Recently, the quantum repeater protocols based on photonic systems have been considered \cite{Chen2017,Xu2017} wherein purification and swapping the entanglement between eight photons have been investigated, experimentally. A new control algorithm and system design for a network of quantum repeater has been presented in \cite{Van2009}, where eight field qubits have been considered in this study. The entanglement based quantum key distribution has been discussed in \cite{Guha2015} by using a linear chain of quantum repeaters employing photon-pair.\\
 The quantum repeater protocol based on entangled coherent states has been already investigated \cite{Sangouard2010}, where in this work, the entangled coherent states have been produced using beam splitter and the produced entanglement has been swapped using linear optics and photon-number resolving detectors. One of the advantages of entangled coherent states as two-mode continuous-variable states \cite{Jeong2001} is that the quasi-Bell states (QBSs) are distinguishable by parity \cite{Munro2000}, wherein heating that changes the vibrational quanta is associated with bit flip errors; while it can be detected and also corrected with the help of appropriate circuit. 
  Entanglement swapping based on QBS measurement (QBSM) has been investigated in \cite{Pakniat2016}. Quantum teleportation of QBS has been considered in \cite{Wang2001,Prakash2007,Li2003,Xin2006}. The QBS can be produced in the presence of Kerr medium \cite{Kuang2007}. Also, in \cite{Jeong2006} weak nonlinearities have been used for the generation of the entangled coherent states. The other way for generation of entangled coherent states is to use the Schr\"{o}dinger cat states as input states of the beam splitter. The Schr\"{o}dinger cat states can be generated by fully quantized picture of the cross phase modulations induced by double-EIT (electromagnetic induced transparency) \cite{Paternostro2003}. In this line, higher-generation of Schr\"{o}dinger cat states in cavity QED has been investigated in \cite{Malbouisson1999}.\\
In this line of research, in our previous work we have considered the quantum repeater based on atomic systems \cite{Ghasemi2018}. In the latter investigation eight two-level atoms $(1,2,\cdots 8)$ were considered where the pairs $(i,i+1)$ with $i=1,3,5,7$, have been prepared in atomic Bell states. Then, by performing interaction between each  two non-entangled adjacent atoms in single- and two-mode cavities, the target atoms $(1,8)$ were finally entangled. Because of the importance of entangled coherent states, we are now motivated to propose a quantum repeater protocol based on the entangled coherent states as two-mode continuous variable state. In fact, in our present work the entanglement between two separate distant locations (A, B) as well as (C, D) is produced by beam splitter where its output state is a QBS with the Schr\"{o}dinger cat state as the input state. Using QBSM method these produced entanglement are swapped to two separable states far distant locations (A, D). Finally, our model is extended to $2^{N}$ locations where $N=3,4,\cdots$; consequently, this model can be used for quantum repeater protocol. To make our model close to real physics, we also consider the lossy effects in our proposed protocol for quantum repeater.\\
\\
 This paper is organized as follows: The entanglement production and entanglement swapping are considered in Sec. 2. The effect of decoherence on swapped entanglement is considered in Sec. 3. Finally, the summary and conclusions are gathered in Sec. 4.
\section{Entanglement production and entanglement swapping}\label{model}

 \begin{figure}[h]
   \centering
 \includegraphics[width=0.45\textwidth]{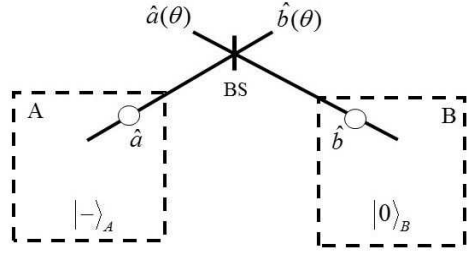}
   \caption{\label{fig:Fig1} {The scheme of entanglement production between two separate distant locations (A, B) using $50:50$ beam splitter. The odd coherent state and vacuum state are input states of beam splitter from locations A and B, respectively.}}
  \end{figure}
\begin{figure}[H]
     \centering
   \includegraphics[width=0.55\textwidth]{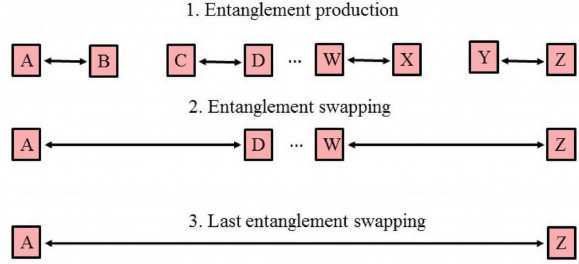}
     \caption{\label{fig:Fig2} {The scheme of quantum repeater protocol: The model can be generalized to $2^N$ separate locations, where $N=3,4,\cdots$, and by performing $2^{N-1}-1$ times of QBSM method, the entanglement is swapped to the first and end locations.}}
    \end{figure}

  In this section we consider entanglement production between two separate adjacent locations A and B. To do this we first introduce a few preliminary and prerequisite concepts and issues. At first, notice that odd coherent state is defined as:
    \begin{eqnarray}\label{oddstate}
  \ket{-}_{\mathrm{A}}&=&\frac{1}{\sqrt{N_-}}(\ket{\alpha}_{\mathrm{A}}-\ket{-\alpha}_{\mathrm{A}})=\frac{e^{-\left|\alpha \right|^2/2 }}{\sqrt{N_-}}\left(e^{\alpha \hat{a}^{\dagger}}-e^{-\alpha \hat{a}^{\dagger}} \right)\ket{0}_{\mathrm{A}}\\ \nonumber
  &=&\frac{1}{\sqrt{N_-}}\left(D_{\hat{a}}(\alpha)-D_{\hat{a}}(-\alpha) \right) \ket{0}_{\mathrm{A}},
    \end{eqnarray}
 where $N_-=2(1-e^{-2\left|\alpha \right|^2 })$ and $\ket{\pm\alpha}_{\mathrm{A}}$ are the standard coherent states, \textit{i.e.}, $\ket{\pm\alpha}=e^{-\left|\alpha \right|^2/2 }\Sigma^\infty_{n=0}\frac{(\pm \alpha)^n}{\sqrt{n!}}\ket{n}$.
 Such states can be generated using different schemes like  propagation in nonlinear medium \cite{Yurke1986,Mecozzi1987,Tombesi1987,Milburn1986},
micromaser cavity experiments \cite{Slosser1990,Wilkens1991} and quantum nondemolition experiments \cite{Song1990}.
  The above odd coherent state and  the vacuum state $\ket{0}_B$ are sent to $50:50$ beam splitter  from separate locations A and B, respectively (see figure \ref{fig:Fig1}). The unitary transformation matrix of this beam splitter has been introduced as \cite{Agarwal2013}:
  \begin{equation}
   U_{\mathrm{BS}}(\theta)= \left( \begin{array}{cc}
  \cos \theta & -\sin \theta  \\
  \sin \theta & \cos \theta  \\ \end{array} \right),
  \label{eq:utmatrix}
  \end{equation}
  where $\cos^2 \theta$ $(\sin^2 \theta)$ are the transmissivity (reflectivity) of the beam splitter. 
  Using of the unitary transformation matrix (\ref{eq:utmatrix}) all the quantum statistical properties of the output fields can be calculated in terms of the properties of the input fields. Also, since the transformation of beam splitter is unitary, there exists an effective Hamiltonian-like for it. Input modes $\hat{a}$ and $\hat{b}$ from locations A and B after passing the beam splitter are combined using unitary transformation (\ref{eq:utmatrix}) as below:
\begin{eqnarray}\label{combinedmodes}
 \hat{a}(\theta)&=&\hat{a}\cos \theta -\hat{b}\sin \theta ,\qquad \hat{b}(\theta)=\hat{a}\sin \theta +\hat{b}\cos \theta .
\end{eqnarray}
It is clear from (\ref{combinedmodes}) that,
\begin{eqnarray}\label{diffcombinedmodes}
 \frac{d}{d\theta}\hat{a}(\theta)=-\hat{b}(\theta),\qquad \frac{d}{d\theta}\hat{b}(\theta)=\hat{a}(\theta).
 \end{eqnarray}
The effective Hamiltonian-like of beam splitter can now be obtained from comparing equation (\ref{diffcombinedmodes}) with Heisenberg-like equation $\frac{d}{d\theta}\hat{c}(\theta)=\frac{1}{i\hbar}\left[\hat{c}(\theta),\hat{H}_{\mathrm{BS}} \right] $ as below:
\begin{eqnarray}\label{ham}
\hat{H}_{\mathrm{BS}}=i \hbar(\hat{a}\hat{b}^\dagger-\hat{a}^\dagger\hat{b}),
 \end{eqnarray}
  where $\hat{c}(\theta)=\hat{a}(\theta),\hat{b}(\theta)$. 
  It should be noticed that the dimension of the eigenvalues of effective Hamiltonian-like $\hat{H}_{\mathrm{BS}}$ is the same as the dimension of  $\hbar$. In fact, in our calculations, as well as in the introduced approach of Ref. \cite{Agarwal2013}, the dimension of eigenvalues of effective Hamiltonian-like does not coincide with energy. 
  Now, using the unitary evolution equation $\hat{u}_{\mathrm{BS}}(\theta)=\exp(-\frac{i}{\hbar}\theta \hat{H}_{\mathrm{BS}})$ with effective Hamiltonian-like (\ref{ham}), the beam splitter operator $\hat{u}_{\mathrm{BS}}(\theta)$ can be obtained as
\begin{eqnarray}\label{matrixbs}
\hat{u}_{\mathrm{BS}}(\theta)=e^{(\hat{a}\hat{b}^\dagger-\hat{a}^\dagger\hat{b})\theta},
 \end{eqnarray}
 by which the following relation is established between equations (\ref{combinedmodes}) and the operator (\ref{matrixbs}) \cite{Gerry2005}:
 \begin{equation}
  \left( \begin{array}{cc}
  \hat{a}(\theta) \\
  \hat{b}(\theta)  \\ \end{array} \right)=\hat{u}^\dagger_{\mathrm{BS}}(\theta)  \left( \begin{array}{cc}
    \hat{a} \\
    \hat{b} \\ \end{array} \right)\hat{u}_{\mathrm{BS}}(\theta).
  \end{equation}
  This makes us sure that the suggested  effective Hamiltonian-like in (\ref{ham}) is introduced correctly.
  Now, let $\ket{-}_{\mathrm{A}}\ket{0}_{\mathrm{B}}$ be the input state, the output state can be obtained using the definition (\ref{oddstate}) and the beam splitter operator (\ref{matrixbs}) as
 \begin{eqnarray}\label{st}
 &&\hat{u}_{\mathrm{BS}}(\theta)\ket{-}_{\mathrm{A}}\ket{0}_{\mathrm{B}}\\\nonumber
 &=&\frac{1}{\sqrt{N_-}}\left(\hat{u}_{\mathrm{BS}}(\theta) D_{\hat{a}}(\alpha)\hat{u}^\dagger_{\mathrm{BS}}(\theta)- \hat{u}_{\mathrm{BS}}(\theta) D_{\hat{a}}(-\alpha)\hat{u}^\dagger_{\mathrm{BS}}(\theta)\right) \hat{u}_{\mathrm{BS}}(\theta)\ket{0}_{\mathrm{A}}\ket{0}_{\mathrm{B}}\\\nonumber
 &=&\frac{1}{\sqrt{N_-}}\left(\ket{\alpha\cos \theta}_{\mathrm{A}}\ket{\alpha\sin \theta}_{\mathrm{B}}-\ket{-\alpha\cos \theta}_{\mathrm{A}}\ket{-\alpha\sin \theta}_{\mathrm{B}}\right).
 \end{eqnarray}
We choose $\theta=\frac{\pi}{4}$ (\textit{i.e.}, the considered beam splitter is $50:50$), so, the resulted state in (\ref{st}) is readily converted into the following QBS \cite{Jeong2002}:
 \begin{eqnarray}\label{qbellab}
 \small
\ket{\varPsi_-}_{\mathrm{AB}}\\\nonumber=\frac{1}{\sqrt{N_-}}\left(\ket{\alpha'}_{\mathrm{A}}\ket{\alpha'}_{\mathrm{B}}-\ket{-\alpha'}_{\mathrm{A}}\ket{-\alpha'}_{\mathrm{B}}\right)\left( =\ket{\mathrm{QBS}}_{\mathrm{AB}}\right),
 \end{eqnarray}
where $\ket{\pm\alpha'}_{\mathrm{A(B)}}=\ket{\pm\frac{\alpha}{\sqrt{2}}}_{\mathrm{A(B)}}=e^{-\left|\alpha' \right|^2/2 }\Sigma^\infty_{n=0}\frac{(\pm \alpha')^n}{\sqrt{n!}}\ket{n}_{\mathrm{A(B)}}$ is the well-known coherent state and $\ket{\mathrm{QBS}}_{\mathrm{AB}}$ is the QBS between A, B. Now, we consider two separate locations C and D shown in figure \ref{fig:Fig2} and the above procedure is repeated for them. Accordingly, the QBS like Eq. (\ref{qbellab}) is obtained for locations C and D as:
 \begin{eqnarray}\label{qbellcd}
\ket{\varPsi_-}_{\mathrm{CD}}\\\nonumber=\frac{1}{\sqrt{N_-}}\left(\ket{\alpha'}_{\mathrm{C}}\ket{\alpha'}_{\mathrm{D}}-\ket{-\alpha'}_{\mathrm{C}}\ket{-\alpha'}_{\mathrm{D}}\right)\left( =\ket{\mathrm{QBS}}_{\mathrm{CD}}\right) .
 \end{eqnarray}
    In the continuation of the paper, the produced entanglement associated with locations (A, B) and (C, D) is swapped to the separate far distant locations (A, D).
 To do this task we use QBSM process. The state of combined locations (A, B, C, D) is $\ket{\Psi}_{\mathrm{AB,CD}}=\ket{\varPsi_-}_{\mathrm{AB}}\otimes\ket{\varPsi_-}_{\mathrm{CD}}$. We operate the QBSM performed with the following QBS \cite{Jeong2002}
\begin{equation}\label{qbell}
\ket{\mathrm{QBS}}_{\mathrm{BC}}=\frac{1}{\sqrt{2(1-e^{-2\left| \beta\right| ^2})}}\left(\ket{\beta}_{\mathrm{B}}\ket{\beta}_{\mathrm{C}}-\ket{-\beta}_{\mathrm{B}}\ket{-\beta}_{\mathrm{C}}\right),
\end{equation}
on the state $\ket{\Psi}_{\mathrm{AB,CD}}$ where $\ket{\pm\beta}_{\mathrm{B}}=e^{-\frac{\left| \beta\right|^2 }{2}}\sum^{\infty}_{n=0}\frac{(\pm\beta)^n}{\sqrt{n!}}\ket{n}_{\mathrm{B}}$. After this operation, the  state in locations (A, D) is converted into the following QBS:
\begin{equation}\label{qbellad}
\ket{\varPsi_-}_{\mathrm{AD}}=\frac{1}{\sqrt{N_-}}\left(\ket{\alpha'}_{\mathrm{A}}\ket{\alpha'}_{\mathrm{D}}-\ket{-\alpha'}_{\mathrm{A}}\ket{-\alpha'}_{\mathrm{D}}\right).
\end{equation}
Without loss of generality $\alpha$ and $\beta$ are assumed to be real. By paying attention to the similarity of QBSs (\ref{qbellab}), (\ref{qbellcd}) and (\ref{qbellad}), our model can be generalized to $2^N$ locations with $N=3,4,\cdots$ where the two end locations are entangled by $2^{N-1}-1$ times operation of QBSM performed with QBS (\ref{qbell}). In the next section, we consider the effect of decoherence on the entanglement swapping.

\section{Effect of decoherence}
\begin{figure}[h]
    \centering
  \includegraphics[width=0.55\textwidth]{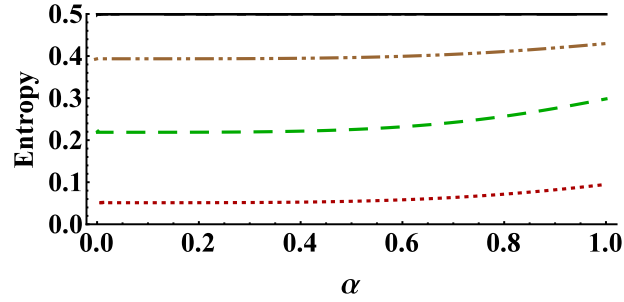}
    \caption{\label{fig:Fig3} {\it The variation of entropy}: for $\eta=0.1$ (dotted red line), $\eta=0.4$ (dashed green line), $\eta=0.7$ (dot-dot-dashed brown line) and $\eta=1$ (solid black line).}
   \end{figure}

\begin{figure}[H]
    \centering
     \subfigure[\label{fig.Fig4a} \ $\beta=\gamma=\alpha'$, $\omega=\mu=-\alpha'$, $\varphi=\pi$]{\includegraphics[width=0.55\textwidth]{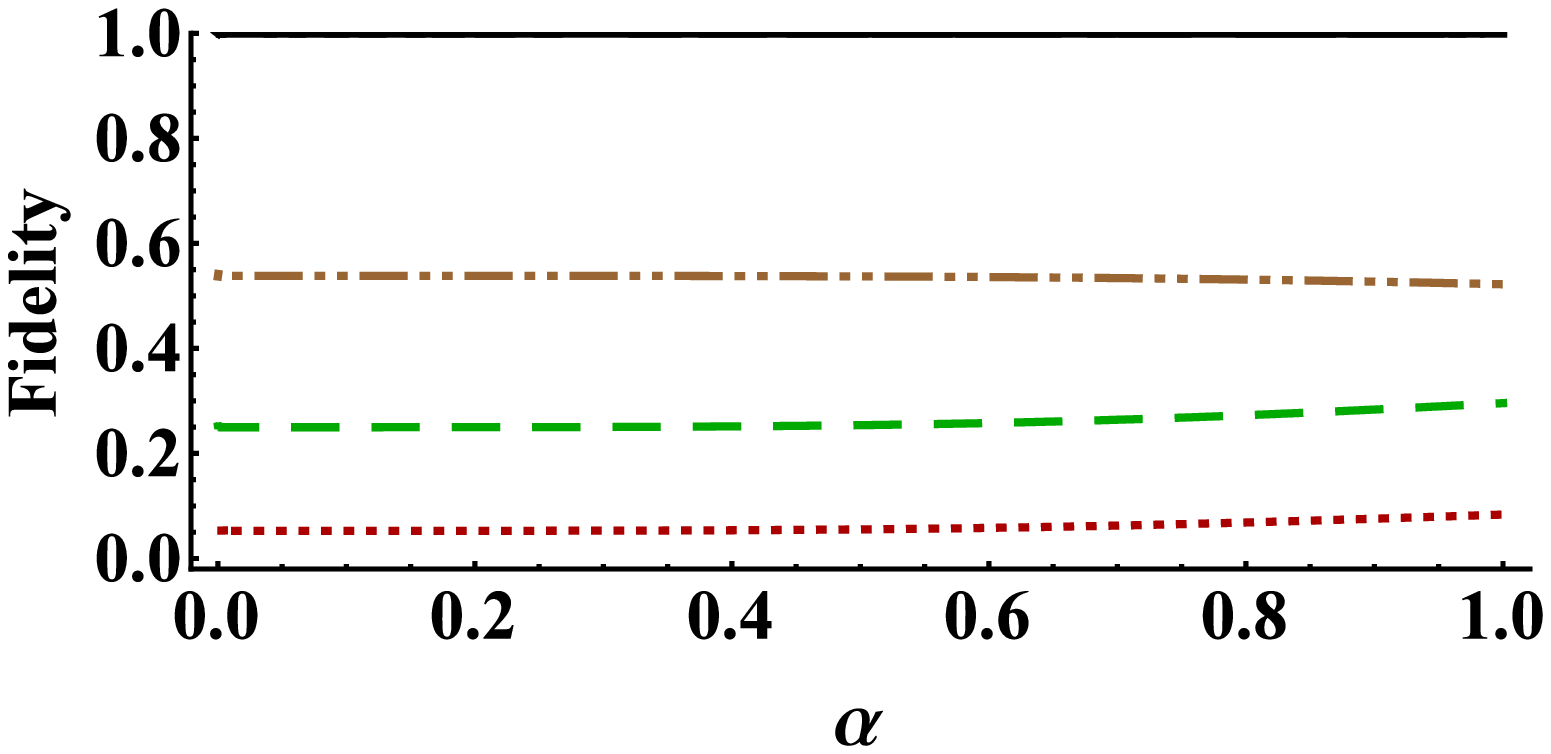}}
        \hspace{0.05\textwidth}
    \subfigure[\label{fig.Fig4b} \ $\beta=\gamma=\alpha'$, $\omega=\mu=-\alpha'$, $\varphi=0$]{\includegraphics[width=0.55\textwidth]{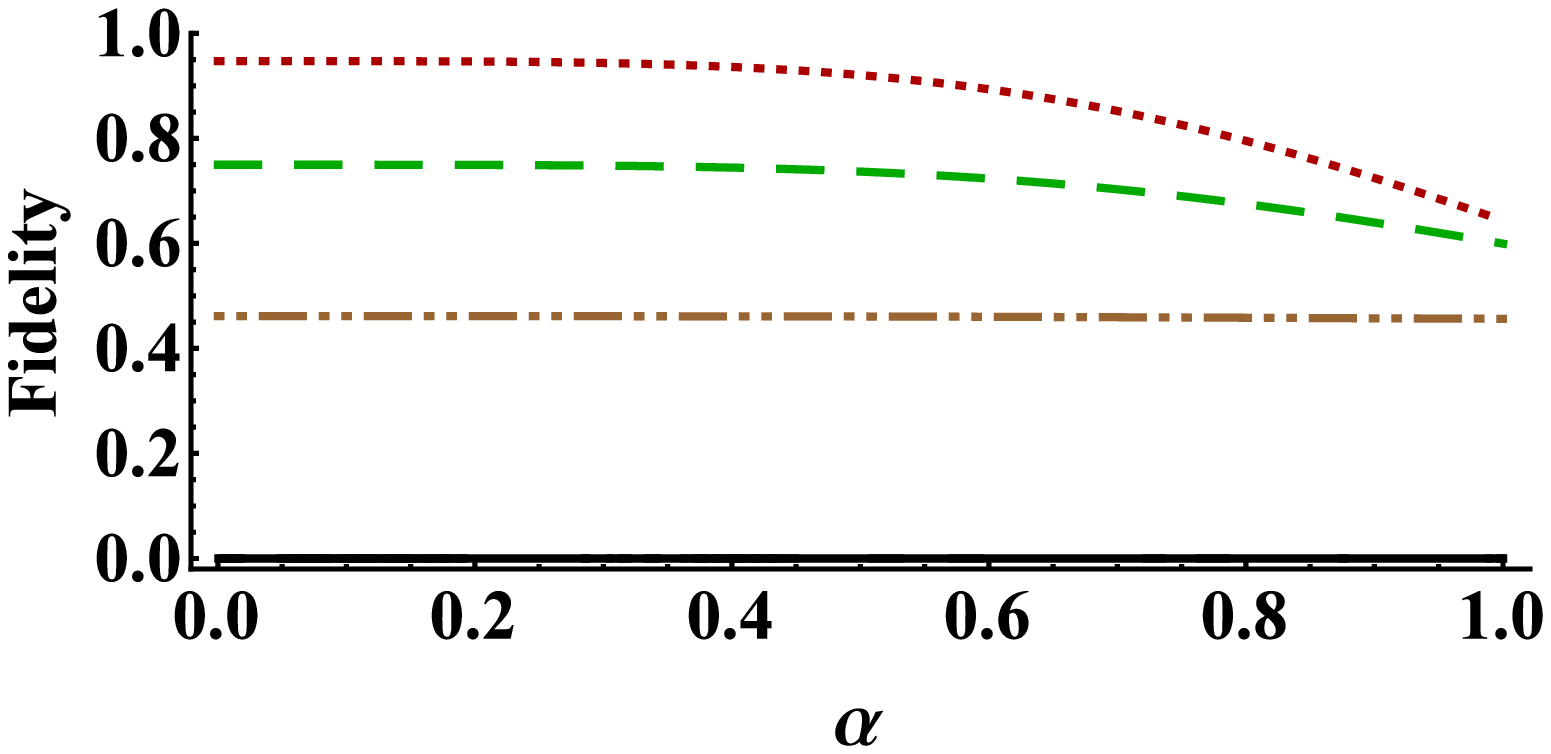}}
    \hspace{0.05\textwidth}
    \subfigure[\label{fig.Fig4c} \ $\beta=\mu=\alpha'$, $\omega=\gamma=-\alpha'$, $\varphi=0$]{\includegraphics[width=0.55\textwidth]{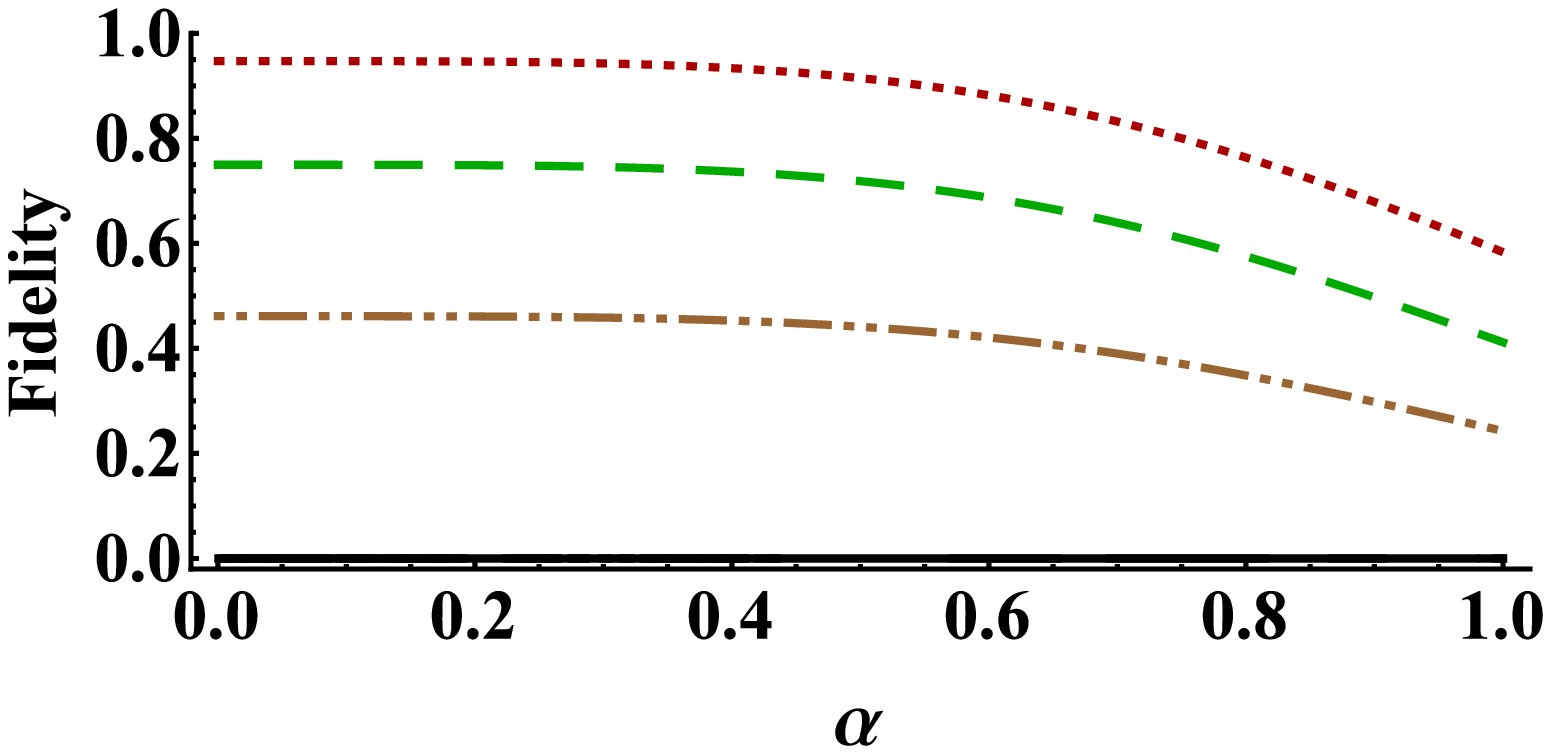}}
    \caption{\label{fig.fid1} {\it The variation of fidelity}: for $\eta=0.1$ (dotted red line), $\eta=0.4$ (dashed green line), $\eta=0.7$ (dot-dot-dashed brown line) and $\eta=1$ (solid black line).}
    \end{figure}

To consider the influence of dissipation,  the produced two-mode entangled states correspond to the  separate locations (A, B) and (C, D) are sent to a noisy channel.
In this respect, the produced QBSs (\ref{qbellab}), (\ref{qbellcd}) are mixed with vacuum states related to the environmental modes. So, the following states
 \begin{eqnarray}\label{qqbellab}
 \small
\ket{\varPsi_-}_{\mathrm{AB}}\otimes\ket{0}_{E_1}\otimes\ket{0}_{E_2},
 \end{eqnarray}
 and
  \begin{eqnarray}\label{qqbellcd}
  \small
 \ket{\varPsi_-}_{\mathrm{CD}}\otimes\ket{0}_{E^{'}_1}\otimes\ket{0}_{E^{'}_2},
  \end{eqnarray}
 are sent into the channel, where the subscripts $\mathrm{E_{1(2)}}$, $\mathrm{E^{'}_{1(2)}}$ relate to the environmental modes.
 The states (\ref{qqbellab}) and (\ref{qqbellcd}), after passing the channel which is characterized by \cite{Bergmann2017,Wickert2010,Van2001}
  \begin{eqnarray}
  \small
 &&\ket{\pm\alpha^{'}}_{A(B)}\otimes\ket{0}_{E_{1(2)}}\rightarrow\ket{\pm\alpha^{'}\sqrt{\eta}}_{A(B)}\otimes\ket{\pm\alpha^{'}\sqrt{1-\eta}}_{E_{1(2)}},\\\nonumber
 &&\ket{\pm\alpha^{'}}_{C(D)}\otimes\ket{0}_{E^{'}_{1(2)}}\rightarrow\ket{\pm\alpha^{'}\sqrt{\eta}}_{C(D)}\otimes\ket{\pm\alpha^{'}\sqrt{1-\eta}}_{E^{'}_{1(2)}},
  \end{eqnarray}
   are converted to the following states, respectively:
 \begin{eqnarray}\label{decqbellab}
&&\ket{\phi_-}_{\mathrm{AB,E_1E_2}}\\\nonumber
&=&\frac{1}{\sqrt{N_-}}\left(\ket{\alpha'\sqrt{\eta}}_{\mathrm{A}}\ket{\alpha'\sqrt{\eta}}_{\mathrm{B}}\ket{\alpha'\sqrt{1-\eta}}_{\mathrm{E_1}}\ket{\alpha'\sqrt{1-\eta}}_{\mathrm{E_2}}\right. \\\nonumber
&-&\left. \ket{-\alpha'\sqrt{\eta}}_{\mathrm{A}}\ket{-\alpha'\sqrt{\eta}}_{\mathrm{B}}\ket{-\alpha'\sqrt{1-\eta}}_{\mathrm{E_1}}\ket{-\alpha'\sqrt{1-\eta}}_{\mathrm{E_2}}\right),
 \end{eqnarray}
 and
  \begin{eqnarray}\label{decqbellcd}
 &&\ket{\phi_-}_{\mathrm{CD,E^{'}_1E^{'}_2}}\\\nonumber
 &=&\frac{1}{\sqrt{N_-}}\left(\ket{\alpha'\sqrt{\eta}}_{\mathrm{C}}\ket{\alpha'\sqrt{\eta}}_{\mathrm{D}}\ket{\alpha'\sqrt{1-\eta}}_{\mathrm{E^{'}_1}}\ket{\alpha'\sqrt{1-\eta}}_{\mathrm{E^{'}_2}}\right. \\\nonumber
 &-&\left. \ket{-\alpha'\sqrt{\eta}}_{\mathrm{C}}\ket{-\alpha'\sqrt{\eta}}_{\mathrm{D}}\ket{-\alpha'\sqrt{1-\eta}}_{\mathrm{E^{'}_1}}\ket{-\alpha'\sqrt{1-\eta}}_{\mathrm{E^{'}_2}}\right),
  \end{eqnarray}
 where $\eta$ is the transmittance of the noisy channel
 \footnote{The photon loss is usually modeled via mixing the light mode with a vacuum state at a beam splitter with transmittance $\eta$ where $1-\eta$ is related to the loss probability of a single-photon. The parameter $\eta$ is defined as $\eta=e^{-\frac{L}{L_{\mathrm{att}}}}$ where $L$ and $L_{\mathrm{att}}$ are optical propagation distance and attenuation length of channel, respectively \cite{Wickert2010,Bergmann2017,Van2001}.}.
  As mentioned above, the parameter $0\leq\eta\leq1$ is related to the transmittance of beam splitter, \textit{i.e.,} for $\eta=0$ (maximum of noise) the entanglement between locations (A, B) as well as (C, D) is washed out, completely.
 Consequently, the entangled state for separate far distant locations (A, D) after operating the QBSM performed with the QBS (\ref{qbell}) on the state $ \ket{\phi_-}_{\mathrm{AB,E_1E_2}} \otimes \ket{\phi_-}_{\mathrm{CD,E^{'}_1E^{'}_2}}$ results in:
 \begin{eqnarray}\label{decqbellad}
  \ket{\phi_-}_{\mathrm{AD,E_1E_2,E^{'}_1E^{'}_2}}=\frac{1}{\sqrt{N^{'}_-}}\left(\ket{\alpha'\sqrt{\eta}}_{\mathrm{A}}\ket{\alpha'\sqrt{\eta}}_{\mathrm{D}}\ket{\alpha'\sqrt{1-\eta}}_{\mathrm{E_1}}\right. \\\nonumber \ket{\alpha'\sqrt{1-\eta}}_{\mathrm{E_2}}  \ket{\alpha'\sqrt{1-\eta}}_{\mathrm{E^{'}_1}}\ket{\alpha'\sqrt{1-\eta}}_{\mathrm{E^{'}_2}}\\\nonumber -  \ket{-\alpha'\sqrt{\eta}}_{\mathrm{A}}\ket{-\alpha'\sqrt{\eta}}_{\mathrm{D}}\ket{-\alpha'\sqrt{1-\eta}}_{\mathrm{E_1}} \ket{-\alpha'\sqrt{1-\eta}}_{\mathrm{E_2}}\\\nonumber\left. \ket{-\alpha'\sqrt{1-\eta}}_{\mathrm{E^{'}_1}}\ket{-\alpha'\sqrt{1-\eta}}_{\mathrm{E^{'}_2}}\right),
  \end{eqnarray}
where $ N^{'}_-=2(1-e^{4(\eta-2)\left|\alpha' \right|^2 })$. The density operator associated with state (\ref{decqbellad}) after tracing over environmental modes can be obtained as:
 \begin{eqnarray}\label{denad}
 \small
\rho_{\mathrm{AD}}&=&\frac{1}{N^{'}_-}\left[ \ket{\alpha'\sqrt{\eta}}_{\mathrm{A}} \bra{\alpha'\sqrt{\eta}}\ket{\alpha'\sqrt{\eta}}_{\mathrm{D}}\bra{\alpha'\sqrt{\eta}}\right. \\\nonumber
&+&\left. \ket{-\alpha'\sqrt{\eta}}_{\mathrm{A}} \bra{-\alpha'\sqrt{\eta}}\ket{-\alpha'\sqrt{\eta}}_{\mathrm{D}}\bra{-\alpha'\sqrt{\eta}}\right. \\\nonumber
 &-&\left. \left( \ket{\alpha'\sqrt{\eta}}_{\mathrm{A}} \bra{-\alpha'\sqrt{\eta}}\ket{\alpha'\sqrt{\eta}}_{\mathrm{D}}\bra{-\alpha'\sqrt{\eta}}\right.\right.  \\\nonumber
 &+&\left. \left. \ket{-\alpha'\sqrt{\eta}}_{\mathrm{A}}  \bra{\alpha'\sqrt{\eta}}\ket{-\alpha'\sqrt{\eta}}_{\mathrm{D}}\bra{\alpha'\sqrt{\eta}}\right) e^{-8(1-\eta)\left|\alpha' \right|^2 }\right].
   \end{eqnarray}
  In this stage, to evaluate the entanglement degree (entanglement between A and (D, environmental modes)) we calculate the entropy \cite{Agarwal2005} corresponding to the above density state via the relation
   \begin{eqnarray}
   \mathrm{S}=1-\mathrm{Tr}_{\mathrm{A}}\rho^2_{\mathrm{A}},
   \end{eqnarray}
   where $\rho_{\mathrm{A}}=\mathrm{Tr}_{\mathrm{D}}\rho_{\mathrm{AD}}$, results in
      \begin{eqnarray}\label{entropy}
     \mathrm{S}&=&1-\frac{2}{N^{'}_-}\left( 1-4e^{-4\left|\alpha \right|^2 }e^{2\eta \left|\alpha \right|^2}+e^{-8\left|\alpha \right|^2}e^{6\eta \left|\alpha \right|^2}\right.  \\\nonumber
     &+&\left.e^{-2\eta \left|\alpha \right|^2}+e^{-8\left|\alpha \right|^2}e^{4\eta \left|\alpha \right|^2}\right) .
      \end{eqnarray}
Also, fidelity of density operator (\ref{denad}) is calculated as:
 \begin{eqnarray}\label{fidelity}
    \mathrm{F}&=&\bra{\Phi}\rho_{\mathrm{AD}}\ket{\Phi}\\\nonumber
    &=&\frac{1}{L N^{'}_-}\left\lbrace e^{-\left|\beta-\sigma \right|^2} e^{-\left|\gamma-\sigma \right|^2}+e^{-\left|\beta+\sigma \right|^2} e^{-\left|\gamma+\sigma \right|^2}\right. \\\nonumber
    &-&\left. 2 e^{-\frac{1}{2}\left|\beta-\sigma\right|^2 } e^{-\frac{1}{2}\left|\beta+\sigma\right|^2 } e^{-\frac{1}{2}\left|\gamma-\sigma \right|^2} e^{-\frac{1}{2}\left|\gamma+\sigma \right|^2}e^{-8(1-\eta)\left|\alpha'\right|^2 }\right. \\\nonumber
    &+&\left. 2 \cos{\varphi} \left[ e^{-\frac{1}{2}\left|\beta-\sigma\right|^2 } e^{-\frac{1}{2}\left|\omega-\sigma\right|^2 } e^{-\frac{1}{2}\left|\gamma-\sigma\right|^2 } e^{-\frac{1}{2}\left|\mu-\sigma\right|^2 }\right. \right. \\\nonumber
    &+&\left. \left. e^{-\frac{1}{2}\left|\beta+\sigma\right|^2 } e^{-\frac{1}{2}\left|\omega+\sigma\right|^2 } e^{-\frac{1}{2}\left|\gamma+\sigma\right|^2 } e^{-\frac{1}{2}\left|\mu+\sigma\right|^2 }\right. \right. \\\nonumber
    &-&\left( e^{-\frac{1}{2}\left|\beta-\sigma\right|^2 } e^{-\frac{1}{2}\left|\omega+\sigma\right|^2 } e^{-\frac{1}{2}\left|\gamma-\sigma\right|^2 } e^{-\frac{1}{2}\left|\mu+\sigma\right|^2 }\right. \\\nonumber
    &+&\left. \left.  e^{-\frac{1}{2}\left|\beta+\sigma\right|^2 } e^{-\frac{1}{2}\left|\omega-\sigma\right|^2 } e^{-\frac{1}{2}\left|\gamma+\sigma\right|^2 } e^{-\frac{1}{2}\left|\mu-\sigma\right|^2 }\right)e^{-8(1-\eta)\left|\alpha'\right|^2}  \right]\\\nonumber
    &+& e^{-\left|\omega-\sigma \right|^2} e^{-\left|\mu-\sigma \right|^2}+e^{-\left|\omega+\sigma \right|^2} e^{-\left|\mu+\sigma \right|^2} \\\nonumber
   &-&\left.  2 e^{-\frac{1}{2}\left|\omega-\sigma\right|^2 } e^{-\frac{1}{2}\left|\omega+\sigma\right|^2 } e^{-\frac{1}{2}\left|\mu-\sigma \right|^2} e^{-\frac{1}{2}\left|\mu+\sigma \right|^2}e^{-8(1-\eta)\left|\alpha'\right|^2} \right\rbrace ,
    \end{eqnarray}
    where
  \begin{eqnarray}
 \ket{\Phi}&=&\frac{1}{\sqrt{L}}\left(\ket{\beta}_{\mathrm{A}}\ket{\gamma}_{\mathrm{D}}+e^{i \varphi}\ket{\omega}_{\mathrm{A}}\ket{\mu}_{\mathrm{D}} \right),\\\nonumber
 L&=&2\left(1+e^{-\frac{1}{2}\left|\beta-\omega\right|^2 } e^{-\frac{1}{2}\left|\gamma-\mu\right|^2 } \cos{\varphi}\right),
   \end{eqnarray}
 where $\sigma=\alpha'\sqrt{\eta}=\alpha\sqrt{\frac{\eta}{2}}$ and $\ket{\beta}$, $\ket{\gamma}$, $\ket{\omega}$ and $\ket{\mu}$ are the standard coherent states, \textit{i.e.}, $\ket{\zeta}=e^{-\left|\zeta \right|^2/2 }\Sigma^\infty_{n=0}\frac{\zeta^n}{\sqrt{n!}}\ket{n}$, where $\zeta=\beta, \gamma, \omega,\mu$.\\
  In this section we have plotted the variation of entropy (\ref{entropy}) (entanglement between A and (D, environmental modes)) as a function of $\alpha$, $\eta$ in figure \ref{fig:Fig3}. From figure \ref{fig:Fig3}, one can see that the entropy has been increased by increasing transparency $\eta$.
  In this figure, state of locations (A, D) is converted to QBS (\ref{qbellad}) for $\eta=1$.\\
 Also, the effects of $\alpha$ and $\eta$ on fidelity (\ref{fidelity}) have been considered in figure \ref{fig.fid1} for different QBSs. In figure \ref{fig.Fig4a}, fidelity has been increased by increasing $\eta$ and for $\eta=1$ the state of locations (A, D) is completely converted to QBS (\ref{qbellad}), but in figures \ref{fig.Fig4b} and \ref{fig.Fig4c} the maxima of fidelity have been decreased by increasing $\eta$. In figures \ref{fig.Fig4b} and \ref{fig.Fig4c} fidelity has been decreased by increasing $\alpha$. Also, fidelity for $\beta=\mu=\alpha'$, $\gamma=\omega=-\alpha'$, $\varphi=\pi$ was calculated and we found that for these considered circumvents, the fidelity is zero.
\section{Summary and conclusions} \label{sec.Conclusion}
In this paper the QBS for two separate distant locations was produced using the $50:50$ beam splitter, where the input states were odd coherent state and vacuum state. Then, the produced entanglement between locations (A, B) and (C, D) was swapped to two separable far distant locations (A, D) using QBSM. Also, the effect of decoherence on the swapped entanglement was considered. The entanglement of the produced entangled states was calculated via entropy showing that the entropy increases by increasing the transparency $\eta$. Also, the maxima of fidelity are decreased by increasing $\eta$ in most cases. We found that because of the stability of produced entangled states, our model can be generalized for $2^N$ separate locations, where $N=3,4,\cdots$ and the entanglement is swapped to two end locations by $2^{N-1}-1$ times of the operation of QBSM. So, we can introduce this model as a quantum repeater protocol to transfer the entangled coherent states to long distances.
   \section*{References}
   \bibliographystyle{iopart-num}
   \bibliography{myreference}

\providecommand{\newblock}{}
\begin{thebibliography}{10}
\expandafter\ifx\csname url\endcsname\relax
  \def\url#1{{\tt #1}}\fi
\expandafter\ifx\csname urlprefix\endcsname\relax\def\urlprefix{URL }\fi
\providecommand{\eprint}[2][]{\url{#2}}

\bibitem{Briegel1998}
Briegel H~J, D{\"u}r W, Cirac J~I and Zoller P 1998 {\em Phys. Rev. Lett.\/}
  {\bf 81} 5932

\bibitem{Duan2001}
Duan L~M, Lukin M, Cirac J~I and Zoller P 2001 {\em Nature\/} {\bf 414} 413

\bibitem{Zhou2015}
Zhou L and Sheng Y~B 2015 {\em Laser Phys. Lett.\/} {\bf 12} 045203

\bibitem{Kaiser2013}
Kaiser F, Issautier A, Ngah L~A, Alibart O, Martin A and Tanzilli S 2013 {\em
  Laser Phys. Lett.\/} {\bf 10} 045202

\bibitem{Agarwal2013}
Agarwal G~S 2012 {\em Quantum optics\/} (Cambridge University Press)

\bibitem{Pakniat2017}
Pakniat R, Zandi M~H and Tavassoly M~K 2017 {\em Eur. Phys. J. Plus\/} {\bf
  132} 3

\bibitem{Kim2002}
Kim M~S, Son W, Bu{\v{z}}ek V and Knight P~L 2002 {\em Phys. Rev. A\/} {\bf 65}
  032323

\bibitem{Berrada2009}
Berrada K, El~Baz M, Saif F, Hassouni Y and Mnia S 2009 {\em J. Phys. A: Math.
  Theor.\/} {\bf 42} 285306

\bibitem{Liao2011}
Liao Q, Fang G, Wang Y, Ahmad M and Liu S 2011 {\em Eur. Phys. J. D\/} {\bf 61}
  475--479

\bibitem{Ghasemi2016}
Ghasemi M and Tavassoly M~K 2016 {\em Eur. Phys. J. Plus\/} {\bf 131} 297

\bibitem{Ghasemi2017}
Ghasemi M, Tavassoly M~K and Nourmandipour A 2017 {\em Eur. Phys. J. Plus\/}
  {\bf 132} 531

\bibitem{Tacsgiotan2010}
Ta{\c{s}}g$\iota$n M, Oktel M, You L and M{\"u}stecapl$\iota$o{\u{g}}lu {\"O}
  2010 {\em Laser Phys.\/} {\bf 20} 700--708

\bibitem{Abdel2008}
Abdel-Khalek S, Khalil E and Ali S 2008 {\em Laser Phys.\/} {\bf 18} 135--143

\bibitem{Bashkirov2005}
Bashkirov E~K 2005 {\em Laser Phys. Lett.\/} {\bf 3} 145

\bibitem{Jiang2009}
Jiang L, Taylor J~M, Nemoto K, Munro W~J, Van~Meter R and Lukin M~D 2009 {\em
  Phys. Rev. A\/} {\bf 79} 032325

\bibitem{Zhao2010}
Zhao B, M{\"u}ller M, Hammerer K and Zoller P 2010 {\em Phys. Rev. A\/} {\bf
  81} 052329

\bibitem{Han2010}
Han Y, He B, Heshami K, Li C~Z and Simon C 2010 {\em Phys. Rev. A\/} {\bf 81}
  052311

\bibitem{Wang2012}
Wang T~J, Song S~Y and Long G~L 2012 {\em Phys. Rev. A\/} {\bf 85} 062311

\bibitem{Wang2014}
Wang C, Wang T and Zhang Y 2014 {\em Laser Phys. Lett.\/} {\bf 11} 065202

\bibitem{Predojevic2015}
Predojevi{\'c} A and Mitchell M~W 2015 {\em Engineering the Atom-Photon
  Interaction\/} (Springer)

\bibitem{Munro2008}
Munro W, Van~Meter R, Louis S~G and Nemoto K 2008 {\em Phys. Rev. Lett.\/} {\bf
  101} 040502

\bibitem{Simon2007}
Simon C, De~Riedmatten H, Afzelius M, Sangouard N, Zbinden H and Gisin N 2007
  {\em Phys. Rev. Lett.\/} {\bf 98} 190503

\bibitem{Chen2017}
Chen L~K, Yong H~L, Xu P, Yao X~C, Xiang T, Li Z~D, Liu C, Lu H, Liu N~L, Li L
  {\em et~al.\/} 2017 {\em Nat. Photon.\/} {\bf 11} 695

\bibitem{Xu2017}
Xu P, Yong H~L, Chen L~K, Liu C, Xiang T, Yao X~C, Lu H, Li Z~D, Liu N~L, Li L
  {\em et~al.\/} 2017 {\em Phys. Rev. Lett.\/} {\bf 119} 170502

\bibitem{Van2009}
Van~Meter R, Ladd T~D, Munro W~J and Nemoto K 2009 {\em IEEE/ACM Trans. Netw.
  (TON)\/} {\bf 17} 1002--1013

\bibitem{Guha2015}
Guha S, Krovi H, Fuchs C~A, Dutton Z, Slater J~A, Simon C and Tittel W 2015
  {\em Phys. Rev. A\/} {\bf 92} 022357

\bibitem{Sangouard2010}
Sangouard N, Simon C, Gisin N, Laurat J, Tualle-Brouri R and Grangier P 2010
  {\em J. Opt. Soc. Am. B\/} {\bf 27} A137--A145

\bibitem{Jeong2001}
Jeong H, Kim M and Lee J 2001 {\em Phys. Rev. A\/} {\bf 64} 052308

\bibitem{Munro2000}
Munro W~J, Milburn G~J and Sanders B~C 2000 {\em Phys. Rev. A\/} {\bf 62}
  052108

\bibitem{Pakniat2016}
Pakniat R, Tavassoly M~K and Zandi M~H 2016 {\em Chin. Phys. B\/} {\bf 25}
  100303

\bibitem{Wang2001}
Wang X 2001 {\em Phys. Rev. A\/} {\bf 64} 022302

\bibitem{Prakash2007}
Prakash H, Chandra N, Prakash R and Shivani 2007 {\em Phys. Rev. A\/} {\bf 75}
  044305

\bibitem{Li2003}
Li S~B and Xu J~B 2003 {\em Phys. Lett. A\/} {\bf 309} 321--328

\bibitem{Xin2006}
Xin-Hua C, Jie-Rong G, Jian-Jun N and Jin-Ping J 2006 {\em Chin. Phys.\/} {\bf
  15} 488

\bibitem{Kuang2007}
Kuang L~M, Chen Z~B and Pan J~W 2007 {\em Phys. Rev. A\/} {\bf 76} 052324

\bibitem{Jeong2006}
Jeong H and An N~B 2006 {\em Phys. Rev. A\/} {\bf 74} 022104

\bibitem{Paternostro2003}
Paternostro M, Kim M and Ham B 2003 {\em Phys. Rev. A\/} {\bf 67} 023811

\bibitem{Malbouisson1999}
Malbouisson J and Baseia B 1999 {\em J. Mod. Opt.\/} {\bf 46} 2015--2041

\bibitem{Ghasemi2018}
Ghasemi M and Tavassoly M~K 2018 {\em EPL (Europhysics Letters)\/} {\bf 123}
  24002

\bibitem{Yurke1986}
Yurke B and Stoler D 1986 {\em Phys. Rev. Lett.\/} {\bf 57} 13

\bibitem{Mecozzi1987}
Mecozzi A and Tombesi P 1987 {\em Phys. Rev. Lett.\/} {\bf 58} 1055

\bibitem{Tombesi1987}
Tombesi P and Mecozzi A 1987 {\em J. Opt. Soc. Am. B\/} {\bf 4} 1700--1709

\bibitem{Milburn1986}
Milburn G~J and Holmes C~A 1986 {\em Phys. Rev. Lett.\/} {\bf 56} 2237

\bibitem{Slosser1990}
Slosser J~J and Meystre P 1990 {\em Phys. Rev. A\/} {\bf 41} 3867

\bibitem{Wilkens1991}
Wilkens M and Meystre P 1991 {\em Phys. Rev. A\/} {\bf 43} 3832

\bibitem{Song1990}
Song S, Caves C~M and Yurke B 1990 {\em Phys. Rev. A\/} {\bf 41} 5261

\bibitem{Gerry2005}
Gerry C and Knight P~L 2005 {\em Introductory quantum optics\/} (Cambridge
  university press)

\bibitem{Jeong2002}
Jeong H and Kim M~S 2002 {\em Phys. Rev. A\/} {\bf 65} 042305

\bibitem{Bergmann2017}
Bergmann M and van Loock P 2019 {\em Phys. Rev. A\/} {\bf 99} 032349

\bibitem{Wickert2010}
Wickert R, Bernardes N~K and van Loock P 2010 {\em Phys. Rev. A\/} {\bf 81}
  062344

\bibitem{Van2001}
van Enk S~J and Hirota O 2001 {\em Phys. Rev. A\/} {\bf 64} 022313

\bibitem{Agarwal2005}
Agarwal G~S and Biswas A 2005 {\em J. Opt. B: Quantum Semiclass. Opt.\/} {\bf
  7} 350

\end{thebibliography}

\end{document}